\documentclass[conference,twoside,final,letterpaper]{IEEEtran}
\usepackage[utf8]{inputenc}
\usepackage[T1]{fontenc}
\usepackage{fixltx2e}
\usepackage{graphicx}
\usepackage{longtable}
\usepackage{float}
\usepackage{wrapfig}
\usepackage{soul}
\usepackage{textcomp}
\usepackage{marvosym}
\usepackage{wasysym}
\usepackage{latexsym}
\usepackage{amssymb}
\usepackage{hyperref}
\usepackage[boxed,noline]{algorithm2e}
\usepackage{multirow}
\IEEEoverridecommandlockouts

\title{\LARGE Toward Fully-Shared Access: Designing ISP Service Plans Leveraging Excess Bandwidth Allocation}
\author{Kyeong Soo Kim\\Department of Electrical and Electronic Engineering\\~Xi'an Jiaotong-Liverpool University\\Suzhou, 215123, P. R. China\\Kyeongsoo.Kim@xjtlu.edu.cn}
\date{Time-stamp: <2014-09-16 12:12:28 Kyeong Soo (Joseph) Kim>}
\hypersetup{
  pdfkeywords={Access, service plan, network pricing, resource allocation, Internet service provider (ISP), traffic shaping, quality of service (QoS)},
  pdfsubject={IEEE CL paper on pricing for fully shared access networks},
  pdfcreator={Emacs Org-mode version 7.9.3f}}

\begin{document}

\maketitle

\begin{abstract}
Shaping subscriber traffic based on token bucket filter (TBF) by Internet
service providers (ISPs) results in waste of network resources in shared access
when there are few active subscribers, because it cannot allocate excess
bandwidth in the long term. New traffic control schemes have been recently
proposed to allocate excess bandwidth among active subscribers proportional to
their token generation rates. In this paper we report the current status of our
research on designing flexible yet practical service plans exploiting excess
bandwidth allocation enabled by the new traffic control schemes in shared access
networks, which are attractive to both ISP and its subscribers in terms of
revenue and quality of service (QoS) and serve as a stepping stone to
fully-shared access in the future.
\end{abstract}

\begin{IEEEkeywords}
Access, service plan, network pricing, resource allocation, Internet service provider (ISP), traffic shaping, quality of service (QoS).
\end{IEEEkeywords}

\pagestyle{empty}       
\thispagestyle{empty}

\section{Introduction}
\label{sec-1}

\begin{figure}[!htb]
    \begin{center}
        \includegraphics[angle=-90,width=\linewidth,trim=15 0 0 0]{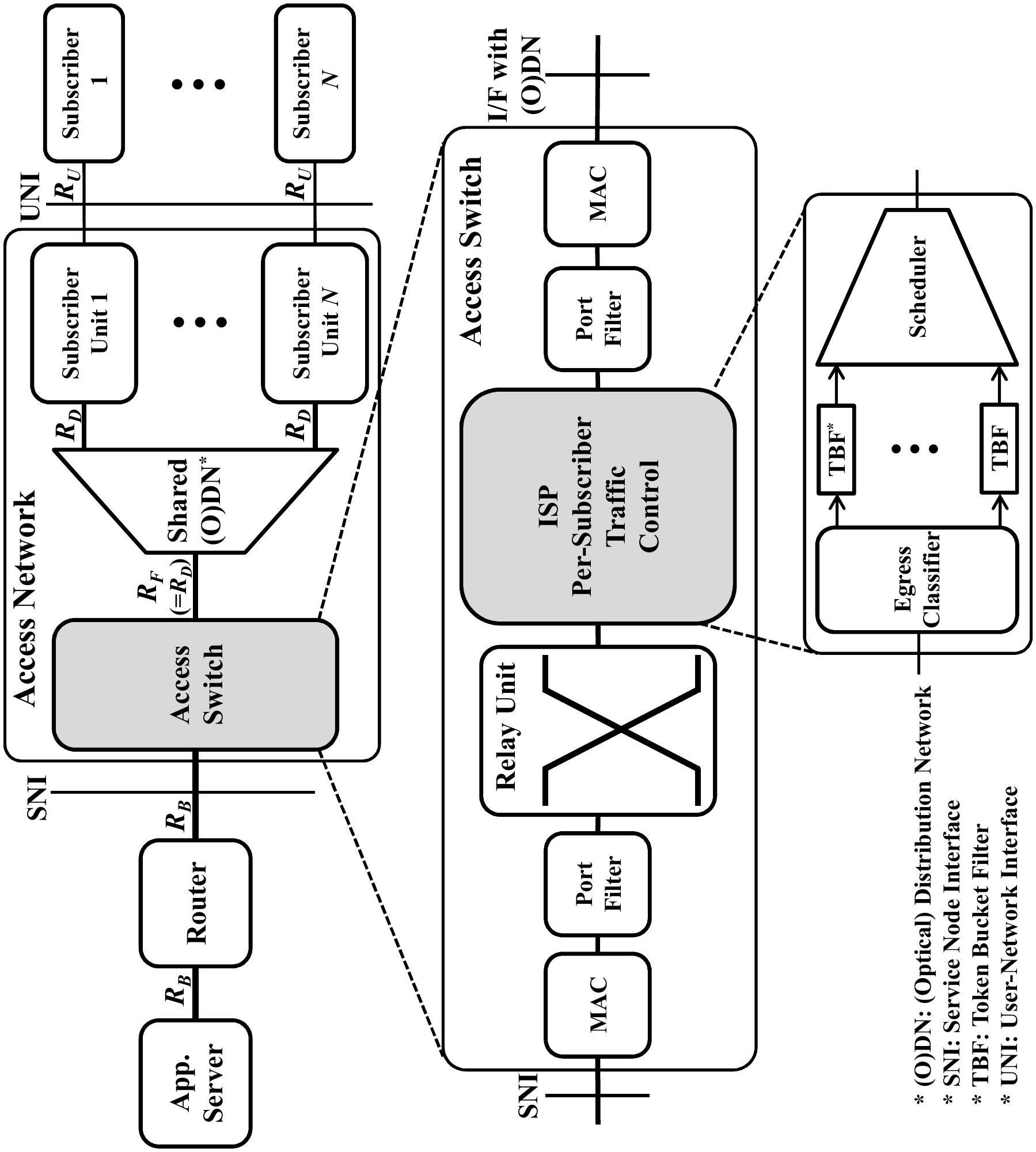}
    \end{center}
\caption{Overview of current practice of ISP traffic control in shared access (shown for downstream traffic only).}
  \label{fg:isp_traffic_control}
\end{figure}

The current implementation and operation of traffic control schemes by Internet
service providers (ISPs) for shared access (e.g., cable Internet or Ethernet
passive optical network (EPON)) prevent subscribers from getting the benefits of
full sharing of resources available in the network; due to the arrangement of
traffic shapers and a scheduler in the access switch shown in
Fig.\(~\)\ref{fg:isp_traffic_control}, the capability of allocating available
bandwidth by the scheduler (e.g., weighted fair queueing (WFQ)) is limited to
\emph{traffic already shaped} by token bucket filters (TBFs) per service contracts
with subscribers. This means that, even though the network is shared access, it
is being managed by ISPs as if it is dedicated access. With the current practice
of ISP traffic control in shared access, therefore, we cannot expect any
fundamental sharing of resources in the long term except for that in the short
term controlled by the size of a token bucket \cite{bauer11:_power,Kim:13-3}.

In fact, the allocation of excess bandwidth in a shared link has long been
discussed in the context of quality of service (QoS) control in routers
(e.g. \cite{HTB}) and recently in the more specific context of ISP traffic
control in shared access in \cite{Kim:14-1} and \cite{Kim:14-3}; the major goal
of the latter discussions is to allocate excess bandwidth among active
subscribers in a fair and efficient way while not compromising the service
contracts specified by the original token bucket algorithm for conformant
subscribers in all time scales and with practical implementation. The business
aspect of the excess bandwidth allocation in ISP traffic control, however, is
yet to be investigated because the flat-rate service plan --- dominant one in
current residential broadband access market --- is so tightly coupled with
traffic shaping based on TBFs that it cannot exploit the excess bandwidth
allocation.

In this paper we report the current status of our research on designing flexible
yet practical ISP service plans leveraging the excess bandwidth allocation
enabled by the recently proposed ISP traffic control schemes in shared access
networks. We first consider an architecture for hybrid ISP traffic control to
gradually introduce the excess bandwidth allocation while providing backward
compatibility with the existing traffic control infrastructure and then discuss
the requirements for a new service plan to provide benefits for both ISP and its
subscribers compared to the existing flat-rate service plans. Then we provide a
design example for a new service plan meeting the requirements derived in this
paper and discuss the impact of key design parameters on ISP revenue and
subscriber QoS.
\section{Designing Service Plans Leveraging Excess Bandwidth Allocation}
\label{sec-2}
\subsection{Hybrid ISP Traffic Control Architecture}
\label{sec-2-1}

Much of existing work on Internet pricing focuses on congestion pricing based on
the game theory (e.g., \cite{cao02:_inter}). As discussed in
\cite{roberts04:_inter_qos}, however, the game-theory-based framework does not
fit the current practice of ISP traffic control and dominant flat-rate service
plans. Rather, a hybrid approach like the proposal for a flexible service plan
based on both flat-rate and usage-based ones in \cite{altmann01:_inter} seems
more practical and appealing to ISPs and subscribers, which is still static in
that there is no interaction between ISP and its subscribers during the
operation.

\begin{figure}[!t]
    \centering
    \includegraphics[angle=-90,width=.9\linewidth,trim=0 0 0 0]{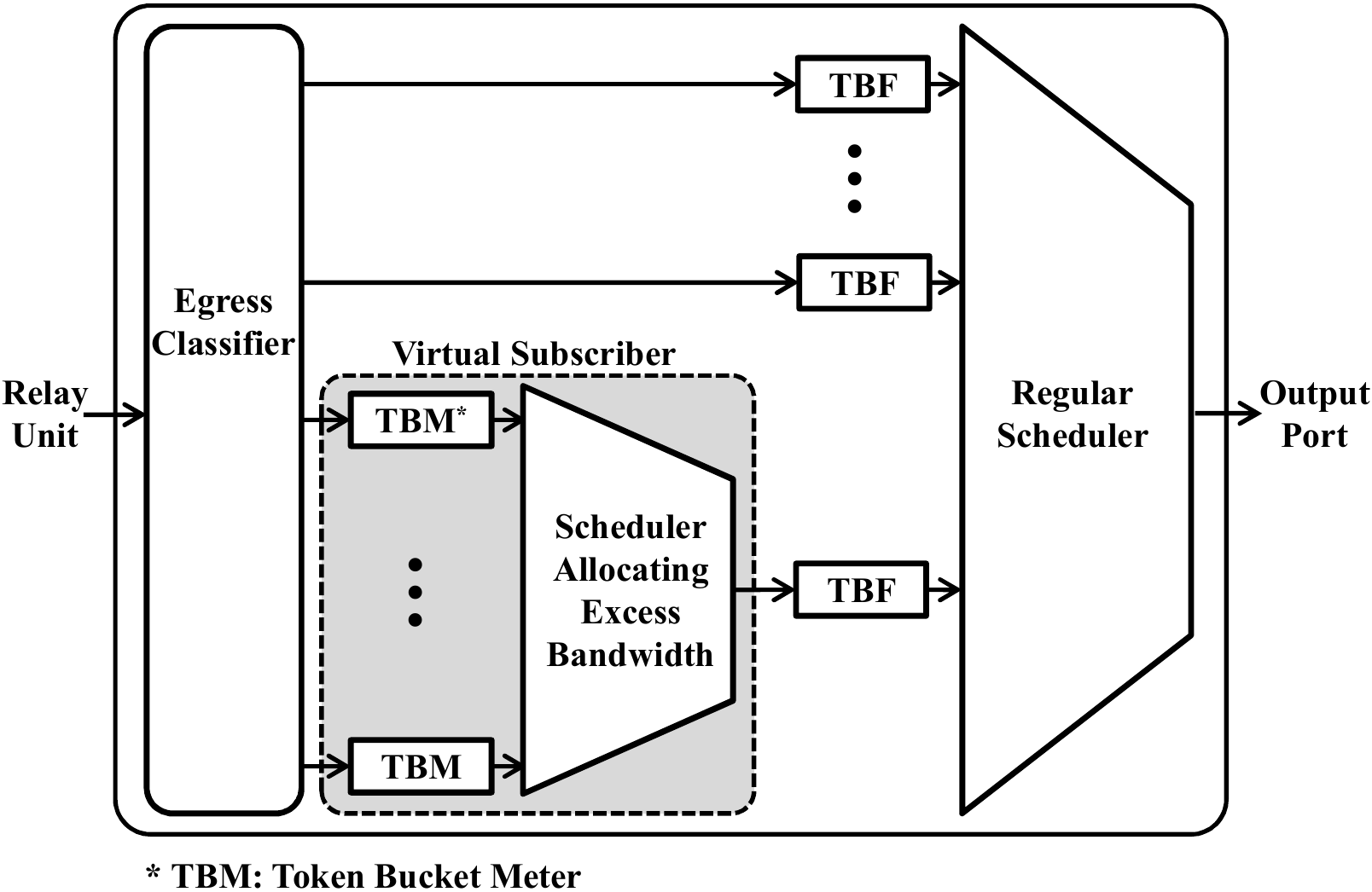}
    \caption{Hybrid ISP traffic control for a flexible service plan
            leveraging excess bandwidth allocation.}
    \label{fg:hybrid_traffic_control}
\end{figure}

Taking the current flat-rate service plan based on traffic shaping by TBFs as a
starting point for our design of a new service plan,
we propose an architecture for hybrid ISP traffic control shown in
Fig.\(~\)\ref{fg:hybrid_traffic_control} in order to gradually introduce the
excess bandwidth allocation in shared access. For backward compatibility with
the existing infrastructure for traffic control and pricing, a group of
subscribers for this new service plan is treated as one \emph{virtual} subscriber
(i.e., their traffic is collectively controlled by a single TBF and also under a
flat-rate service plan during the plan design procedure); for a new service plan
leveraging excess bandwidth allocation, on the other hand, the traffic from each
subscriber belonging to this group is individually controlled by the new ISP
traffic control scheme enabling excess bandwidth allocation within the group.

Note that the migration toward fully-shared access will be finished when those
subscribers under traditional TBF-based traffic shaping and flat-rate service
plans also move under new traffic control schemes enabling excess bandwidth
allocation and new flexible service plans, where all the network resources are
fully and efficiently shared among the subscribers.
\subsection{Requirements \& Design Guidelines for New Service Plans}
\label{sec-2-2}

For a new service plan to be acceptable, it is desirable to guarantee that there
should be no disadvantage for both ISP and its subscribers compared to the
existing flat-rate service plans.
Here we discuss and derive requirements for a new service plan in terms of
parameters for existing flat-rate service plans.

Consider the case of designing a new service plan based on the two existing
flat-rate service plans specified as follows:
\begin{itemize}
\item Lower-rate flat-rate service plan (for the individual subscribers)
\begin{itemize}
\item Monthly price: $P_{L}$
\item Token generation rate: $TGR_{L}$
\item Token bucket size: $TBS_{L}$
\end{itemize}
\item Higher-rate flat-rate service plan (for the virtual subscriber)
\begin{itemize}
\item Monthly price: $P_{H}$
\item Token generation rate: $TGR_{H}$
\item Token bucket size: $TBS_{H}$
\end{itemize}
\end{itemize}

Based on the arrangement of traffic shaper and new traffic control scheme
enabling excess bandwidth allocation shown in
Fig.\(~\)\ref{fg:hybrid_traffic_control}, we are to provide a new service plan
which is specified as follows:
\begin{itemize}
\item New hybrid service plan
\begin{itemize}
\item Number of subscribers: $N$
\item Monthly price: $P+P(u)$
\item Token generation rate: $TGR_{L}$
\item Token bucket size: $TBS_{L}$
\end{itemize}
\end{itemize}
$P$ is a fixed, minimum price, and $P(u)$ is a usage-based price function where
$u$ is the usage of excess bandwidth (i.e., the amount of non-conformant
traffic). Note that this group of $N$ subscribers is considered as one virtual
subscriber as for traffic management and pricing in deriving requirements for a
new service plan in the following.

First of all, for this new service plan to be feasible in terms of subscriber
QoS, the sum of token generation rates for lower-rate plan subscribers should be
no greater than the token generation rate of higher-rate plan, i.e.,
\begin{equation}
    N \times TGR_L \leq TGR_H ,
\end{equation}
or
\begin{equation}
    N \leq \frac{TGR_H}{TGR_L} .
    \label{eq:N_condition1}
\end{equation}

From ISP's perspective, the revenue from the new service plans should be no less
than either of that from the higher-rate flat-rate service plan for the virtual
subscriber or the sum of those from the lower-rate flat-rate service plan for
individual subscribers, i.e.,
\begingroup
    \setlength{\arraycolsep}{0.0em}
    \begin{eqnarray}
    \sum_{i=1}^{N}\left\{P+P(u_{i})\right\} &{\geq}& P_{H}, \label{eq:isp_requirement1} \\
\sum_{i=1}^{N}\left\{P+P(u_{i})\right\} &{\geq}& \sum_{i=1}^{N}P_{L}, \label{eq:isp_requirement2}
    \end{eqnarray}
\endgroup
where $u_{i}$ is the usage of excess bandwidth by the \(i\)th subscriber.

From subscribers' perspective, on the other hand, the price for the new service
plan should be no greater than that of the higher-rate flat-rate service plan
and, when there is no usage of excess bandwidth, equal to that of the lower-rate
one, i.e.,
\begingroup
    \setlength{\arraycolsep}{0.0em}
    \begin{eqnarray}
        P+P(u_{max}) &{\leq}& P_{H}, \label{eq:subscriber_requirement1} \\
        P+P(0) &{=}& P_{L}, \label{eq:subscriber_requirement2}
    \end{eqnarray}
\endgroup
where $u_{max}$ is the maximum amount of excess bandwidth that one subscriber
can use for a month. $u_{max}$ corresponds to the extreme case where only one
subscriber uses all the available excess bandwidth for a whole month with all
other subscribers inactive, i.e.,
\begin{equation}
    u_{max}=\left(TGR_{H}-TGR_{L}\right)T_{month} + \left(TBS_{H} - TBS_{L}\right), 
    \label{eq:u_max}
\end{equation}
where $T_{month}$ is a time period for a month. Considering the first term is
usually much larger than the second term in (\ref{eq:u_max}), $u_{max}$ can be
approximated as follows:
\begin{equation}
    u_{max} \approx (TGR_{H}-TGR_{L})T_{month} .
    \label{eq:u_max_approx}
\end{equation}

As for $P(u)$, we assume that it is a monotone-increasing function with
$P(0)=0$, which gives $P=P_{L}$ from (\ref{eq:subscriber_requirement2}). In case
of a simple linear function, i.e., $P(u)=\alpha u$ for a nonnegative constant
$\alpha$, we obtain the following from (\ref{eq:isp_requirement1}) and
(\ref{eq:subscriber_requirement1}) using (\ref{eq:u_max_approx}):
\begingroup
    \setlength{\arraycolsep}{0.0em}
    \begin{eqnarray}
    \alpha\sum_{i=1}^{N} u_i &{\geq}& P_{H}-N \times P_{L}, \label{eq:alpha1} \\
    \alpha &{\leq}& \frac{P_{H}-P_{L}}{\left(TGR_{H}-TGR_{L}\right)T_{month}}. \label{eq:alpha2}
    \end{eqnarray}
In (\ref{eq:alpha1}), because $\sum_{i=1}^{N} u_i$ is bounded by $u_{max}$, it
can be rewritten as follows:
\begin{equation}
    \alpha \geq \frac{P_{H}-N \times P_{L}}{\left(TGR_{H}-TGR_{L}\right)T_{month}}. \label{eq:alpha3}
\end{equation}
Also, because the left-hand side of (\ref{eq:alpha1}) becomes zero when $u_i =
0$ for $\forall i$, the right-hand side should be no greater than zero, i.e.,
\begin{equation}
    P_{H} - N \times P_{L} \leq 0, \label{eq:flat_rate_prices}
\end{equation}
or
\begin{equation}
    N \geq \frac{P_{H}}{P_{L}}, \label{eq:N_condition2}
\end{equation}

Note that, given the two existing flat-rate service plans and a simple linear
usage-based price function, (\ref{eq:N_condition1}) \& (\ref{eq:N_condition2})
and (\ref{eq:alpha2}) \& (\ref{eq:alpha3}) provide requirements for the the slope
of linear price function (i.e., $\alpha$) and the number of subscribers (i.e.,
$N$), respectively. The guidelines for designing a new hybrid service plan based
on the results in this section are summarized in Table
\ref{tbl:design_guidelines}.
\begin{table*}[!t]
  \begin{center}
    \caption{Design guidelines for a new hybrid service plan}
    \label{tbl:design_guidelines}
    \begin{tabular}{|l|l|l|l|}
        \hline
        \multicolumn{1}{|c|}{\multirow{2}{*}{Parameters}} & \multicolumn{2}{|c|}{Flat-rate service plans} & \multicolumn{1}{|c|}{\multirow{2}{*}{Hybrid service plan}} \\ \cline{2-3}
        & \multicolumn{1}{|c|}{Lower rate} & \multicolumn{1}{|c|}{Higher rate} & \\ \hline\hline
        Token generation rate & $TGR_L$ & $TGR_H$ & $TGR_L$ \\ \hline
        \multirow{3}{*}{Monthly price} & \multirow{3}{*}{$P_L$} & \multirow{3}{*}{$P_H$} & $P_L + \alpha u$ \\
        & & & where $\alpha \geq \max\left(0,~\frac{P_{H}-N \times P_{L}}{\left(TGR_{H}-TGR_{L}\right)T_{month}}\right)$ \\
        & & & and $\alpha \leq \frac{P_{H}-P_{L}}{\left(TGR_{H}-TGR_{L}\right)T_{month}}$. \\ \hline
        \multirow{2}{*}{Number of subscribers} & & & $N$ \\
        & & & where $\frac{P_H}{P_L} \leq N \leq \frac{TGR_H}{TGR_L}$. \\ \hline
    \end{tabular}
  \end{center}
\end{table*}
\subsection{A Design Example}
\label{sec-2-3}

To illustrate how to design a new service plan using the requirements and design
guidelines described in Sec. \ref{sec-2-2}, here we provide a design example based on the flat-rate service plans
from Virgin Media Cable Internet \cite{virgin_media_broadband}.

According to their service plans, the lowest tier provides up to 50 Mbit/s speed
for \pounds26.50 per month, while the highest tier up to 152 Mbit/s speed for
\pounds39 per month. The flat-rate service plan parameters, therefore, can be
summarized as follows:
\begin{itemize}
\item Lower-rate flat-rate service plan
\begin{itemize}
\item Monthly price ($P_{L}$): \pounds26.50 per month
\item Token generation rate ($TGR_{L}$): 50 Mbit/s
\end{itemize}
\item Higher-rate flat-rate service plan
\begin{itemize}
\item Monthly price ($P_{H}$): \pounds39 per month
\item Token generation rate ($TGR_{H}$): 152 Mbit/s
\end{itemize}
\end{itemize}
Note that token bucket sizes for both service plans are unknown, but they are
not used in the requirements for design parameters as discussed in \ref{sec-2-2}. $T_{month}$ (assuming 30 days per
month) and $u_{max}$ in this case become 2.592e+6 s and 2.644e+8 Mbit,
respectively.

From Table \ref{tbl:design_guidelines}, we obtain
\begin{equation}
    \frac{39}{26.5} \approx 1.472 \leq N \leq \frac{152}{50} = 3.04.
\end{equation}
For this design example, we set $N$ to 3, i.e., the maximum possible value. For
the slope of linear usage-based price function, we obtain
\begin{equation}
    0 \leq \alpha \leq 4.728\mbox{e-}8~\mbox{\pounds/Mbit} .
\end{equation}
If we take a maximum value for $\alpha$ to maximize the revenue of ISP, the
resulting hybrid service plan is specified as follows:
\begin{itemize}
\item Hybrid service plan
\begin{itemize}
\item Number of subscribers: 3
\item Monthly price: \pounds$\left(26.50 + 4.728\mbox{e-}8 \times u\right)$
\item Token generation rate: 50 Mbit/s
\end{itemize}
\end{itemize}
where $u$ is the usage of excess bandwidth in Mbit.

With this new hybrid service plan, we consider two extreme cases of excess
bandwidth utilization:

First, consider a case where only one subscriber is active and utilizes all the
available excess bandwidth for the whole month (i.e., \(u\)=$u_{max}$). The
revenue is \pounds92, which is higher than either $P_H$ or three times of $P_L$.

Second, consider another case where all three subscribers are active and evenly
divide the available excess bandwidth (i.e.,
\(u\)=$\frac{152-3\times50}{3}\times2.592\mbox{e+}6$ Mbit). The revenue in this
case is \pounds79.745, which is still higher than either $P_H$ or three times of
$P_L$.

This design example demonstrates that the new hybrid service plan enables
subscribers to enjoy benefits of both flat-rate service plans through excess
bandwidth allocation without much increase in monthly payments, while improving
the revenue of ISP depending on the usage patterns of subscribers.
\section{Concluding Remarks}
\label{sec-3}

In this paper we have discussed the issues in current practice of ISP traffic
shaping and related flat-rate service plans in shared access networks and
suggested alternative service plans based on new ISP traffic control schemes
enabling excess bandwidth. We have proposed a hybrid ISP traffic control
architecture to gradually introduce the excess bandwidth allocation in shared
access, and, based on the proposed architecture, provided requirements for new
service plans leveraging excess bandwidth allocation.

Note that the impact of aggregate traffic from the group of subscribers under a
new service plan on metro and backbone networks is yet to be investigated
because it is likely that the utilization of a group of subscribers is higher
than that of a single subscriber under the higher-rate flat-rate service plan.



\end{document}